# Active Learning-Driven Lightweight YOLOv9: Enhancing Efficiency in Smart Agriculture

Hung-Chih Tu[1]*, Bo-Syun Chen[2], Yun-Chien Cheng[1]*,

[1] Department of Mechanical Engineering, College of Engineering, National Yang Ming Chiao Tung University, Hsin-Chu, Taiwan

[2] Industrial Technology Research Institute, Taiwan

*Corresponding author: hungchih.en12@nycu.edu.tw, yccheng@nycu.edu.tw

**Abstract**

This study addresses the demand for real-time detection of tomatoes and tomato flowers by agricultural robots deployed on edge devices in greenhouse environments. Under practical imaging conditions, object detection systems often face challenges such as large scale variations caused by varying camera distances, severe occlusion from plant structures, and highly imbalanced class distributions. These factors make conventional object detection approaches that rely on fully annotated datasets difficult to simultaneously achieve high detection accuracy and deployment efficiency. To overcome these limitations, this research proposes an active learning–driven lightweight object detection framework, integrating data analysis, model design, and training strategy. First, the size distribution of objects in raw agricultural images is analyzed to redefine an operational target range, thereby improving learning stability under real-world conditions. Second, an efficient feature extraction module is incorporated to reduce computational cost, while a lightweight attention mechanism is introduced to enhance feature representation under multi-scale and occluded scenarios. Finally, an active learning strategy is employed to iteratively select high-information samples for annotation and training under a limited labeling budget, effectively improving the recognition performance of minority and small-object categories. Experimental results demonstrate that, while maintaining a low parameter count and inference cost suitable for edge-device deployment, the proposed method effectively improves the detection performance of tomatoes and tomato flowers in raw images. Under limited annotation conditions, the framework achieves an overall detection accuracy of 67.8% mAP, validating its practicality and feasibility for intelligent agricultural applications.

## 1. Introduction

Agriculture has traditionally relied on intensive manual labor[1]. With the development of mechanized and smart agriculture, robots are increasingly used to improve productivity and reduce labor costs. However, for delicate crops such as tomatoes, conventional large machinery is unsuitable because fruits bruise easily and ripen non-uniformly, making it difficult to define a single harvesting standard. In greenhouse environments, these challenges are compounded by complex plant structures and limited labor availability, creating a strong demand for robotic systems that can perform low-damage, real-time harvesting and monitoring[2].

Tomato production further depends on reliable flower pollination. Although tomatoes are self-pollinating, greenhouse conditions lack natural wind and insects, so pollination often relies on manual operations or robotic assistance. Both harvesting and pollination require precise detection and localization of tomatoes and tomato flowers under varying growth stages and appearances. A unified perception system that can robustly detect both objects in raw greenhouse imagery is therefore crucial for improving yield and quality.

Deep learning–based object detection models have been widely explored for tomato-related tasks, including leaf disease identification, plant-height estimation, and flower detection[3-7]. These studies demonstrate the potential of convolutional neural networks for agricultural perception, but most are developed under controlled imaging conditions—close-range views, simple backgrounds, and stable illumination. Models and datasets are usually constructed on standardized images with relatively balanced class distributions, and rarely account for severe occlusion, large scale variation, or extreme class imbalance, especially the scarcity of tomato-flower samples compared with tomatoes or green tomatoes. As a result, directly applying these methods to raw greenhouse video frames with cluttered backgrounds often leads to unstable and biased detection performance.

For deployment on agricultural robots and embedded edge devices, object detection models must also satisfy stringent constraints on computation, memory, and power. Recent work has proposed lightweight detection networks using pruning, quantization, and efficient architectural redesign to reduce parameters and FLOPs while retaining acceptable accuracy[8-10]. However, most approaches focus on simplifying a single structural level (e.g., backbone or neck) and seldom jointly consider multi-scale feature preservation, key-region emphasis, and the learning difficulty induced by data imbalance. In scenes dominated by small objects, background noise, and occlusion, insufficient feature representation can cause critical targets—such as small flowers partially occluded by leaves—to be suppressed in deep feature maps.

To enhance feature selectivity under limited resources, attention mechanisms have been introduced into lightweight detectors. Channel–attention modules such as ECA and SE improve feature weighting with modest computational overhead[11, 12], yet mainly focus on channel relationships and provide limited modeling of spatial structure. In contrast, Pyramid Split Attention (PSA) jointly exploits channel and spatial information through multi-scale feature splitting and fusion, offering stronger potential in complex backgrounds with multi-scale targets[13]. However, PSA also increases computational cost, and its integration into a truly lightweight architecture must be carefully designed.

In agricultural imagery, the scale distribution of objects is highly non-uniform: uncropped greenhouse frames may simultaneously contain large, close-up fruits and very small, distant flowers or immature fruits. Multi-scale feature fusion networks (e.g., FPN-based designs) and data-level strategies such as cropping and mosaic augmentation have been proposed to mitigate this issue[14-16]. Cropping-based methods can increase the relative size of small objects but may discard global context, while image-stitching methods such as Mosaic and Mosaic9 retain larger context and diversify scale distribution at relatively low cost. Their actual benefit, however, still depends on the underlying feature-extraction capacity and the statistics of the training data, particularly under class-imbalanced and heavily occluded conditions.

Another major bottleneck in agricultural perception is annotation cost. Fully labeling large-scale raw image streams is extremely time-consuming, especially when many objects are small, occluded, or blurred. Active Learning (AL) has therefore been explored to iteratively select the most informative samples for annotation, reducing labeling effort while improving training efficiency[17-19]. Yet most existing AL frameworks implicitly assume limited occlusion and relatively balanced classes; their sample-selection stability and computational overhead under greenhouse conditions with small-object dominance and severe class imbalance remain insufficiently studied.

Motivated by these limitations, this study targets real greenhouse videos containing tomatoes, green tomatoes, and tomato flowers, with the goal of developing an object detection framework that is both lightweight and annotation efficient for edge deployment. We build upon the YOLOv9 architecture and redesign the feature-extraction backbone with C3Ghost-based modules, integrate a PSA-style lightweight attention mechanism to enhance multi-scale and spatial awareness, introduce a dynamic Mosaic9 augmentation schedule to strengthen small-object learning without harming convergence, and incorporate a confidence-based active learning strategy tailored to the characteristics of agricultural imagery.

The main contributions of this work are as follows:

1. Develop a CAD-based object detection system designed specifically for real-scene images to address the lack of specialized models.

2. Optimize the model architecture with efficient architecture design and lightweight attention method to enable real-time inference on resource-limited robotic systems.

3. Introducing augmentation methods to support model regain multi-scale features to increase model detect performance.

4. Incorporate active learning to reduce annotation costs and enhance adaptability to continuously increasing real-world data.

In summary, this study is motivated by practical challenges encountered in real agricultural environments, and focuses on addressing limitations arising from raw image inputs, multi-scale variability, and class imbalance. By exploring feasible research directions that balance model performance and data utilization efficiency, this work aims to provide concrete and practical solutions for the deployment of agricultural object detection systems in real-world applications.

## 2. Related Work

### 2.1. YOLOv9

YOLOv9 is the latest generation of the YOLO (You Only Look Once) series of real-time object detection models[20]. The model architecture is shown in Figure 1. Its architecture builds upon the advantages of previous YOLO frameworks, further enhancing feature extraction capabilities and detection accuracy. With a more efficient network design, YOLOv9 achieves faster inference speed and superior object detection performance.

The YOLOv9 architecture consists of three main components: backbone, neck, and head. First, the model utilizes a Convolutional Neural Network (CNN) to extract features from images and incorporates Positional Encoding to preserve spatial information. These features are then processed through a Feature Pyramid Network (FPN) and a Path Aggregation Network (PAN) to enhance multi-scale feature learning, improving accuracy in detecting small objects.

In YOLOv9, the Detection Head adopts a decoupled structure, separating the classification (Cls) and bounding box regression (Reg) tasks. This separation minimizes feature interference and improves detection performance. Additionally, the model employs a dynamic anchor mechanism during training to adapt to objects of different categories and scales, enhancing generalization ability.

Compared to other deep learning architectures, YOLOv9 strikes an optimal balance between inference speed and accuracy, making it particularly suitable for real-time applications such as autonomous driving, intelligent surveillance, and robotic vision systems.

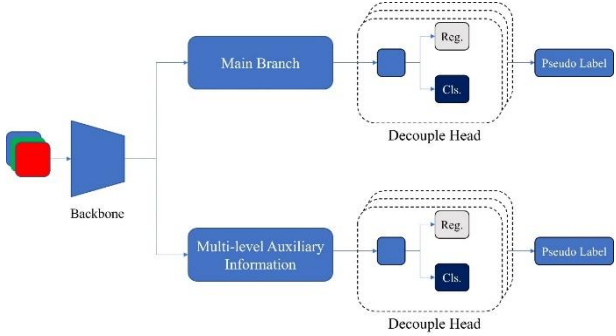

Figure 1 YOLOv9 model architecture

## 2.2. Model lightweight Method

Model lightweight is a technique used to reduce the number of parameters or improve inference speed, making it suitable for resource-constrained applications such as edge devices and mobile devices. Various approaches are used to achieve lightweight models, including Efficient Network Design, Network Pruning, and Knowledge Distillation, which help enhance detection performance under computational constraints.

However, YOLOv9's RepNCSPELAN4 feature extraction module has been shown to require a large number of parameters, leading to high computational resource consumption. This is particularly restrictive for small object detection tasks. Therefore, a key challenge in improving the model is reducing computational overhead while maintaining accuracy.

Among the various lightweight techniques, efficient network design is one of the most commonly adopted strategies. Chen et al. proposed a module that incorporates C3Ghost[21], as shown in Figure 2, to replace the original RepNCSPELAN4 module in YOLOv9, serving as the primary approach for a lightweight backbone.

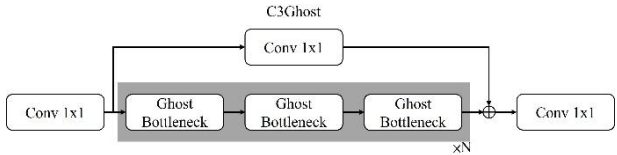

Figure 2 C3Ghost architecture

## 2.3. Lightweight Attention Method

In deep learning-based object detection models, attention mechanisms enhance feature extraction capabilities, allowing the model to focus more effectively on target regions. However, traditional attention mechanisms, such as Transformer, typically require high computational costs, making them difficult to deploy on resource-constrained edge devices. To address this issue, researchers have proposed lightweight attention mechanisms, aiming to reduce computational overhead while still improving feature learning capabilities.

A common approach in attention mechanisms is to combine global, channel, and spatial information to enhance feature representation. However, this often comes with high computational costs. Some lightweight methods address this by using 1D convolution to reduce the complexity introduced by attention mechanisms.

A more efficient approach is Pyramid Split Attention (PSA), which integrates the advantages of both multi-scale feature fusion and lightweight design. This method improves the recognition of objects of varying sizes. The Khanam research team introduced the C2PSA module in YOLOv11[22], incorporating the Pyramid Split Attention (PSA) mechanism. This mechanism is designed to optimize the model's effectiveness in multi-scale feature processing, enhance feature extraction capabilities, and improve the precision and robustness of object detection.

## 2.4. Mosaic9

In object detection tasks, data augmentation strategies are widely employed to enhance a model's ability to adapt to variations in object scale, position, and background. This is particularly important in agricultural imaging, where raw images are typically uncropped, and a single frame often contains both mature fruits and very small flowers or young fruits. As a result, the model is exposed to a highly non-uniform scale distribution during training. Therefore, how to improve the model's capacity to learn multi-scale targets without significantly increasing computational overhead has become a key issue in data-level design.

In application scenarios involving high-resolution raw images and a very high proportion of small objects, Xiao et al. proposed the Mosaic9 augmentation strategy in autonomous driving research[16]. By stitching nine images into a single training sample, the model is exposed to a richer variety of scale variations and target combinations within each iteration, thereby further strengthening its multi-scale feature learning capability.

## 2.5. Active Learning

Active learning is an iterative training process that selects and annotates the most informative unlabeled images to improve model performance. The entire process consists of four main steps[19], as shown in Fig3:

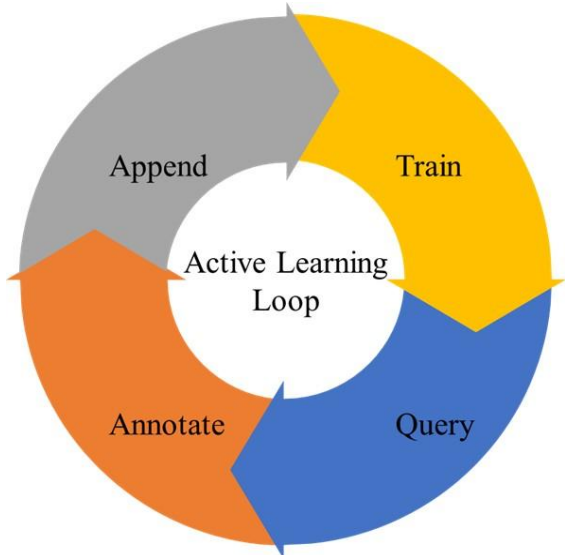

Figure 3. AL flowchart

1. TRAIN: Train the model on an initially labeled dataset. This step establishes a baseline model using the available labeled data.

2. QUERY: Use an acquisition function to select the most informative or representative samples from the unlabeled dataset. These samples are typically the ones where the model has the highest uncertainty or has the greatest potential to improve model performance.

3. ANNOTATE: Human experts label the selected samples. This step is crucial because the correctness of annotations depends on expert knowledge, ensuring high-quality training data.

4. APPEND: The newly labeled samples are added to the training dataset. As more samples get labeled, the model progressively learns from a richer dataset, improving its accuracy and robustness.

Recent studies have integrated this approach into YOLOv7, demonstrating promising performance on public datasets[19, 23]. These results highlight the strong adaptability and potential of active learning for enhancing object detection models.

### 3. Material and Methods

**3.1. Tomato Dataset**

The images used in this study were sourced from videos collected by personnel from the Industrial Technology Research Institute (ITRI) in a greenhouse where tomatoes are grown. A total of 95 video clips were obtained, from which images were extracted at 0.5-second intervals, each image is in RGB format, and the dataset was randomly split into training and validation sets. The training set contains 2186 annotated images, the validation set contains 578 annotated images.

Table 1. Dataset images and annotations quantities

|       | Image | Annotation |              |               |
|-------|-------|------------|--------------|---------------|
| Total | 2734  | Tomato     | Green Tomato | Tomato Flower |
| Train | 2186  | 7424       | 2607         | 1091          |
| Valid | 548   | 1872       | 620          | 278           |

**3.2. Experimental Procedure**

The experimental workflow of this study is illustrated in Figure 4. First, only a small subset of images from the full dataset is manually annotated to serve as the initial training data, which is then divided into a training set and a validation set, simulating real-world scenarios where annotation budgets are limited. At this stage, the model can only learn from a small amount of labeled data and thus serves as the baseline model for the subsequent active learning process.

Based on the initial data distribution and the characteristics of agricultural images—namely the prevalence of multi-scale targets and a high proportion of small objects—this study redesigns the model architecture using YOLOv9 as the backbone. Lightweight feature extraction modules and attention mechanisms are introduced to reduce the number of parameters and computational complexity, while enhancing the model's ability to recognize critical regions and small targets. In parallel, data augmentation strategies are adjusted to improve training stability under limited data conditions.

Once the lightweight architecture has been established and demonstrates stable baseline performance, an active learning mechanism is further integrated. Unlabeled data are evaluated according to sample uncertainty, and representative images are progressively selected for manual annotation. Newly labeled samples are then added back into the training set to update the model. This loop is repeated to assess the practical benefits of active learning in improving model performance while reducing overall annotation cost.

Finally, this study conducts several tests and evaluations, including:
(1) Comparison of different image augmentation methods.
(2) Analysis of the impact of various lightweight techniques on the model.
(3) Comparison of dataset size at each stage under different active learning strategies.
(4) Comparison of test results across multiple models.

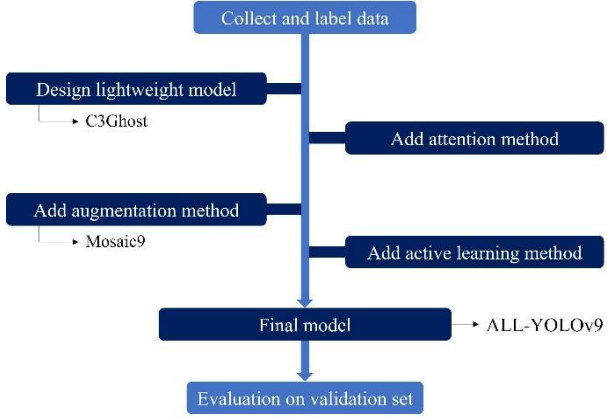

Figure 4. Experiment process

### 3.3. Model Overview

The model designed in this study, ALL-YOLOv9, as shown in Figure 6, is an architecture based on YOLOv9. It integrates an Active Learning module to iteratively improve training data annotation and a lightweight LWA-YOLO to enhance computational efficiency while maintaining detection accuracy. The following sections provide a detailed explanation of these components.

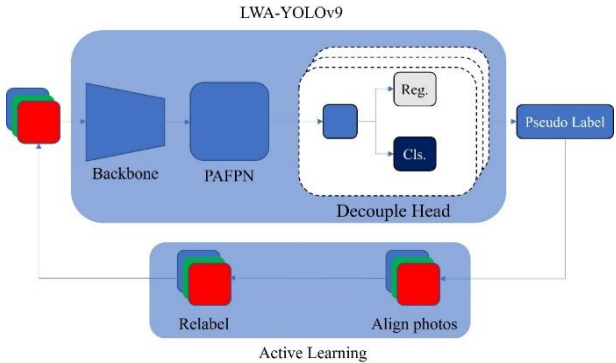

Figure 5. Overview of ALL-YOLOv9

### 3.3.1. LWA(Lightweight Attention)-YOLO

The module adopted in this study is shown in Figure 7. The C3Ghost[21] module combines the Ghost Module and Ghost Bottleneck[24] within a YOLO-style C3/CSP split-and-merge structure. The input feature map is first split into two paths: one path adjusts the channel dimension through a convolution layer and is then directly preserved; the other path is processed by a 1×1 convolution followed by three stacked Ghost Bottlenecks. Finally, the two paths are concatenated and fused by another 1×1 convolution to produce the output.

Each Ghost Bottleneck is composed of two serially connected Ghost Modules with a residual addition between input and output. In a Ghost Module, core features are first generated by a pointwise convolution, and additional features are then synthesized using computationally cheaper depthwise convolutions. These are concatenated to achieve feature expansion and extraction with reduced computational cost and fewer parameters.

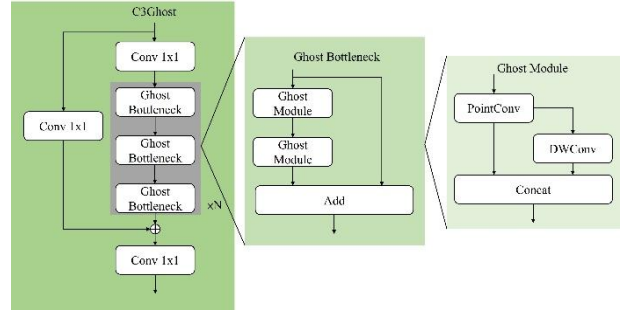

Figure 6. C3Ghost architecture

The C2PSA[22] module is built upon the C2f mechanism, as illustrated in Figure 9, embedding a multi-head attention mechanism within its structure. The attention mechanism follows the equations below:

$$Attention(Q,K,V) = softmax\left(\frac{QK^T}{\sqrt{d_k}}\right)V$$

$$softmax(x)_i = \frac{exp\,(x_i)}{\sum_j exp\,(x_j)}$$

Thus, the C2PSA module is formed, combining channel and spatial information to provide more effective feature extraction. It optimizes the feature maps from the previous layer and enriches them with attention mechanisms to enhance model performance. PSA (Pyramid Split Attention) is an efficient local self-attention module that processes a portion of the convolved features through a structure composed of Multi-Head Self-Attention (MHSA) and a Feed-Forward Network (FFN). The two feature parts are then concatenated and fused through convolution. This approach enhances global modeling capability while reducing computational complexity, improving both accuracy and efficiency.

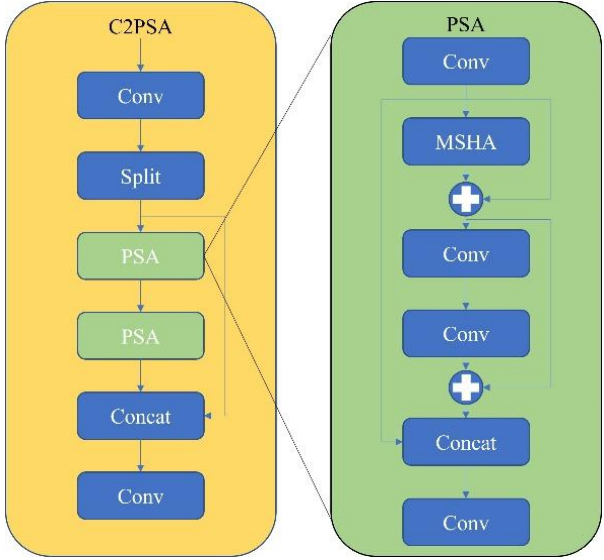

Figure 8. C2PSA architecture

### 3.3.2. Dynamic Mosaic9

To enhance the model's ability to recognize multi-scale and small objects in complex agricultural scenarios, this study introduces the Mosaic9[16] data augmentation strategy during training. Compared with the traditional Mosaic4, which stitches only four images, Mosaic9 further combines nine different images, allowing a single training sample to present richer variations in object scale and spatial distribution.

However, excessive use of Mosaic9 may cause a mismatch between the training image distribution and the real inference scenario. This issue is particularly critical in the later stages of training: once the model has already acquired basic feature extraction capability, overly strong data perturbations may adversely affect convergence stability.

Therefore, this study adopts a dynamic Mosaic on–off mechanism, disabling Mosaic9 in the later training phase so that the model can gradually adapt to the original image distribution and improve its generalization ability in the final inference stage. The switching condition is defined as follows:

$$M(e) = \begin{cases} 1, & e < E - C \\ 0, & e \geq E - C \end{cases}$$

where $e$ denotes the current training epoch, $E$ is the total number of training epochs, and $C$ is the epoch threshold for disabling Mosaic. When $M(e)=1$, Mosaic9 data augmentation is enabled; when $M(e)=0$, only original images are used for training. Through this strategy, Mosaic9 can be fully exploited in the early phase to enrich data diversity, while the later phase gradually returns to the real image distribution, balancing learning efficiency and inference stability.

### 3.3.3. Active learning

In addition to the model architecture design, this study also addresses the common issues of high annotation cost and class imbalance in agricultural images by introducing an Active Learning (AL) mechanism to improve the utilization efficiency of labeled data. The overall active learning pipeline, as illustrated in Figure 3, consists of four iterative stages: model training, sample querying, manual annotation, and dataset expansion.

1. Uncertainty-based sample selection strategy:

Among various active learning methods, entropy-based measures and Monte Carlo Dropout can effectively quantify model uncertainty, but their computational cost is relatively high, which is unfavorable for real-time and resource-constrained agricultural applications. Therefore, this study adopts a confidence-based uncertainty measure, Least Confidence (LC), as the primary criterion for sample selection, in order to balance computational efficiency with practical deployment feasibility. For the $i$-th object predicted in a single image, its uncertainty score is defined as:

$$u_i(x) = 1 - s_i$$

where $s_i$ denotes the confidence score of the predicted class. A lower confidence score indicates higher uncertainty in the model's prediction.

2. Uncertainty aggregation strategies:

Since a single image may contain multiple predicted bounding boxes, this study further aggregates object-level uncertainty scores into a single image-level indicator to facilitate sample ranking and selection. Considering that different aggregation strategies may lead to distinct sample selection behaviors, we compare three image-level uncertainty computation methods:

Average: $U_{average}(x) = \frac{1}{N_x}\sum_{i=1}^{N_x} u_i(x)$

Max: $U_{max}(x) = \max_{1 \leq i \leq N_x} u_x(x)$

Sum: $U_{sum}(x) = \sum_{i=1}^{N_x} u_x(x)$

Here, $N_x$ denotes the number of detected objects (bounding boxes) in image $x$.

The rationale for choosing these three methods is that the intrinsic characteristics of our dataset are not yet fully understood. We therefore aim to investigate whether the dataset exhibits more prominent patterns in terms of object-wise feature difficulty, extreme uncertainty, or object quantity per image.

3. Image Sampling Strategy:

By applying the above aggregation strategies, the model can rank images according to their overall uncertainty and select the top-$K$ images (as defined in this study) for annotation in the next round. The image selection strategy is defined as follows:

$$U_{best}^{(t)} = Top - K\{argmax_{x \in U} U(x)\}, K = 500$$

$$Move\ Strategy: U^{(t+1)} = U^{(t)} \backslash U_{best}^{(t)}$$

$$Copy\ Strategy: U^{(t+1)} = U^{(t)}$$

Through the active learning mechanism, the model can preferentially acquire image samples with higher information content, thereby reducing the annotation cost associated with redundant or low-contribution data. Under a limited annotation budget, this process progressively enhances the model's ability to detect small objects and minority classes in raw agricultural images.

The choice of $K$=500 in each iteration is based on the practical consideration that, in the early deployment phase, it is desirable to improve model performance as rapidly as possible. Compared with selecting only a small number of highly specific samples, selecting a relatively larger batch of images per round provides more substantial training signals, leading to faster performance gains.

**3.4. Data Annotations Preprocessing and Analysis**

To verify the characteristics of object-scale distribution in the dataset used in this study and to provide a reasonable basis for the design of subsequent data preprocessing strategies, we first conduct a systematic statistical analysis of the bounding-box sizes of annotated objects, drawing on dataset analysis methods commonly adopted in small object detection research. Compared with defining small objects using a fixed pixel size, recent studies have pointed out that, under varying image resolutions and scene conditions, using the ratio of bounding-box area to the entire image area (area ratio) as the criterion can more accurately reflect how perceptible an object is within an image.

In the field of small object detection, studies have indicated that when the area of a bounding box is less than approximately 0.58% of the total image area, the object often suffers from insufficient resolution and semantic information attenuation in deep learning models, making it prone to disappearing in deeper feature maps and thereby degrading recognition performance[25]. Consequently, this range is commonly used as an important reference for determining whether a dataset can be regarded as small-object–oriented.

Based on the above definition, this study uses the COCO dataset as a reference and adopts the corresponding area-ratio threshold for small objects in COCO, approximately 0.58% of the image area, to analyze the size distribution of all annotated objects in our dataset. As shown in Figure 9, most annotated objects fall below this threshold, indicating that the dataset in this study exhibits a pronounced small-object characteristic in terms of scale distribution, consistent with the criteria used in small object detection research.

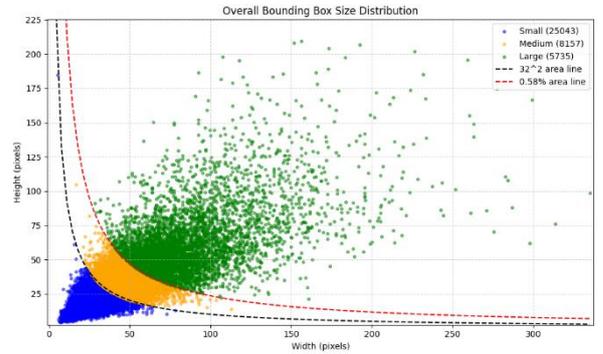

Figure 9. Original dataset object scale diagram

However, further examination of the data distribution reveals that, after the input images are resized to a fixed resolution for model training, the bounding boxes of extremely small objects shrink significantly with the scaling of the image. In many cases, such objects appear at distances so far that they are barely recognizable to the human eye and hold limited practical value in real agricultural robot operations. In addition, in the deeper feature maps of YOLO-series models (such as the P4 and P5 layers), these extremely small objects are often difficult to preserve effectively due to limited feature resolution, and may instead introduce noise or interference into the learning process.

Building on the above analysis, this study does not arbitrarily remove small-object annotations. Instead, following dataset scale–characterization practices in the small object detection literature, we further filter out only extremely small objects and retain annotations for medium- and larger-scale targets. This ensures that the model concentrates on objects that are visually discernible and practically relevant in real agricultural settings, thereby improving overall training stability and detection performance. The size distribution of the screened items is shown in Figure 10.

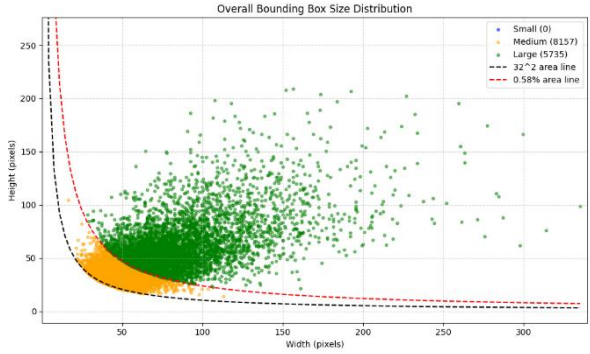

Figure 10 Retaining the scale diagram of medium and large object datasets

### 3.5. Equipment

This study utilized a server with ASUS Z790-A GAMING WIFI 6E, powered by Intel I9-13900K, and equipped with MSI RTX 4090 GAMING X TRIO 24G.

## 4. Result

### 4.1. Evaluation Methods

For all object detection models in this study, we evaluate performance using precision (P), recall (R), and mean average precision (mAP). Average precision (AP) is computed from the precision–recall curve obtained by ranking predictions by confidence and comparing them with ground-truth labels; mAP is the mean AP over the three classes (tomato, green tomato, and tomato flower). In view of deployment on agricultural robots and edge devices, we additionally report the number of parameters and GFLOPs as indicators of memory footprint and computational cost, respectively, with all values computed under the same input resolution for fair comparison. As no public tomato object-detection benchmark is available, all experiments are conducted on the dataset provided by the National Industrial Research Institute.

### 4.2. Implement Detail

The hyperparameters used in training ALL-YOLOv9 in this study include an initial learning rate of 0.01, the optimizer is Adaptive Moment Estimation (Adam), the classification loss is Binary Cross Entropy Loss, and the bounding box loss is a combination of CIoU Loss and Distribution Focal Loss. The total training epoch for ALL-YOLOv9 is 500, the learning rate was reduced to 0.01 times the original value at the end of training.

### 4.3. Ablation on Lightweight Modules

This section compares the impact of different efficient module designs on model performance, and the results are summarized in Table 2. First, when the original RepNCSPELAN4 module in YOLOv9 is replaced solely with the C3Ghost module, the number of parameters and FLOPs can be effectively reduced. However, because the Ghost Bottleneck mainly generates features through linear operations, its ability to represent multi-scale objects and complex backgrounds in deeper feature maps (such as the P4 and P5 layers) is clearly limited, leading to a significant drop in overall mAP.

When the C2PSA attention module is further introduced, the model's capability for channel- and spatial-level feature reweighting is indeed enhanced. Nevertheless, under the condition that the backbone network does not provide sufficiently rich deep features, the attention mechanism cannot effectively mine discriminative key information. As a result, the overall detection performance fails to improve and even exhibits a slight decline.

These findings indicate that attention mechanisms alone cannot compensate for the deficiency in feature representation caused by excessive lightweight; their effectiveness strongly depends on the quality of the underlying features and the completeness of multi-scale information. This also suggests that, in multi-scale and highly complex scenarios such as agricultural imagery, purely structural lightweight or relying solely on attention as compensation is insufficient to balance model efficiency and recognition accuracy. It is necessary to further incorporate designs at the data level or training-strategy level to fully realize overall performance gains.

Table 2. Model Ablation Study

| | C3Ghost | PSA | Param. | Flops | $mAP_{50}$ |
|---|---|---|---|---|---|
| LWA-YOLOv9 | No | No | 2.66M | 10.7G | 54.1 |
| | Yes | No | 1.67M | 7.2G | 47.9 |
| | Yes | Yes | 1.77M | 7.3G | 46.8 |

### 4.4. Effect of Dynamic Mosaic9

In agricultural image-based object detection tasks, object scales are often unevenly distributed, with a high proportion of small objects, making the model highly sensitive to scale variations during training. To enhance multi-scale feature learning, YOLO-based models commonly employ Mosaic data augmentation, which stitches multiple images together to increase scale diversity. However, prolonged use of Mosaic may cause the model, especially in the later training stages, to overfit to the stitched image distribution, thereby degrading its ability to recognize objects at their original scales.

Based on these considerations, this study further investigates the impact of different Mosaic configurations on model performance, including the number of stitched images and the timing of disabling Mosaic. To enrich the scale diversity of objects in training samples, the original Mosaic setting of four-image stitching is extended to nine images, thereby enhancing the representation of object scales. At the same time, considering that inference in real agricultural scenarios is still performed on original, non-stitched images, Mosaic augmentation is disabled in the later training phase. Specifically, the Mosaic usage ratio is set to 80% of the total training epochs, allowing the model to readapt to the original scale distribution of images during convergence.

As shown in Table 3, when using nine-image Mosaic and disabling this augmentation in the final stage of training, the mAP of the LWA-YOLOv9 model improves from 46.8% (without disabling Mosaic) to 48.9%. These results indicate that appropriately controlling the usage schedule of Mosaic helps balance multi-scale feature learning and adaptation to the original image distribution, thereby enhancing detection performance in real agricultural imaging scenarios.

Table 3. Detection Accuracy Comparison During Active Learning Stages

| Mosaic9 | 80%Epoch On | Model | $mAP_{50}$ |
|---|---|---|---|
| No | No | YOLOv9t | 54.1 |
| | | Our model(no PSA) | 47.9 |
| | | LWA-YOLOv9(Ours) | 46.8 |
| Yes | No | YOLOv9t | 54.1 |
| | | Our model (no PSA) | 47.9 |
| | | LWA-YOLOv9(Ours) | 46.8 |
| No | Yes | YOLOv9t | 53 |
| | | Our model (no PSA) | 47.3 |
| | | LWA-YOLOv9(Ours) | 46.8 |
| Yes | Yes | YOLOv9t | 54.1 |
| | | Our model (no PSA) | 47.3 |
| | | LWA-YOLOv9(Ours) | 48.9 |

**4.5. Active Learning Performance**

This section compares active learning sample selection strategies and their effects on annotation growth and class distribution. We examine three uncertainty aggregation methods—Average, Max, and Sum—to identify which more effectively selects informative images in multi-object, class-imbalanced agricultural scenes.

After each training round, the model's bounding-box predictions on unlabeled images are used to compute an image-level uncertainty score. Images are then ranked by this score and selected for annotation. Because each image may contain multiple targets, the aggregation rule shapes the selection behavior: the Average method uses the mean uncertainty of all boxes; the Max method uses the highest uncertainty; and the Sum method accumulates the uncertainties of all boxes to reflect the overall difficulty of the image. Appendix A summarizes how the number of newly annotated images evolves under these strategies. In most stages, Max and Sum select more images than Average, indicating a stronger tendency to target highly uncertain or densely annotated images.

The class statistics in Appendix B and C show that these differences become more pronounced under class imbalance. For Average, the image score is easily dominated by many low-uncertainty predictions, so the contribution of truly difficult or rare cases (e.g., small or heavily occluded objects) is diluted, reducing their chance of being selected. By contrast, Max focuses directly on the hardest prediction in each image, whereas Sum captures the cumulative learning value of multiple medium- to high-uncertainty objects within a single image.

In terms of class distribution, the tomato class exhibits the largest increase in annotations under Max and Sum, while growth for green tomatoes and tomato flowers is more limited. This reflects the co-occurrence pattern in the dataset: mature tomatoes frequently appear alongside other categories, so strategies driven by peak or accumulated uncertainty tend to repeatedly select images containing tomatoes.

Overall, in the raw agricultural image setting considered here, Max and Sum are more effective than Average at mining high-information samples, especially in multi-object, class-imbalanced scenarios with many small objects. These results support the choice of Max and Sum in the proposed active learning framework and underpin subsequent gains in detection performance.

**4.6. Active learning comparison**

Although each active learning strategy was constrained to select 500 images per round to ensure fairness in data quantity, Appendix D shows that different sample selection policies still yield clear performance gaps. This confirms that the quality of selected samples, rather than quantity alone, is critical for effective model learning.

Under the move mechanism, comparing the three uncertainty aggregation methods (Average, Max, and Sum) shows that Max achieves the highest mAP in most stages, with particularly clear gains between the 4th and 6th rounds. Using the most uncertain prediction in each image as the selection criterion allows the model to focus on samples it currently finds most difficult.

Comparing the move and copy data-handling strategies further reveals a key difference in sample accumulation. In the copy strategy, selected images are added to the training set but remain in the unlabeled pool, so the same images may be selected repeatedly, limiting the growth of unique annotated samples and reducing overall information gain. In contrast, the move strategy removes selected high-uncertainty images from the unlabeled pool and adds them to the labeled set, ensuring that each round introduces genuinely new, informative samples. This leads to more pronounced performance improvements in early rounds and is better suited to scenarios with limited annotation budgets.

Notably, with the move strategy, the Max method reaches performance comparable to the fully labeled setting as early as the 4th round, where mAP improves from 67.1 to 67.8. This suggests that, given the characteristics of our dataset—each image often containing multiple targets, where a single highly uncertain object can make the entire image valuable—the Max strategy is particularly effective at identifying informative samples.

In summary, under the settings of this study, combining the move data strategy with Max-based uncertainty aggregation most effectively improves both the practical utility and learning efficiency of the proposed agricultural object detection framework.

**4.7. Class imbalance validation**

As shown in Appendix E, the dataset exhibits a clear annotation imbalance among the three classes (tomato, green tomato, and tomato flower): tomatoes have the most annotations, whereas tomato flowers form a minority class. This imbalance strongly influences both the training dynamics and the final recognition performance.

In the initial dataset (230 images), the tomato flower class has only 110 annotations, yielding an mAP of 21.1, substantially lower than that of tomatoes and green tomatoes. This indicates that, under severe data scarcity, the model cannot effectively learn discriminative features for this class. After introducing the active learning strategy, high-uncertainty samples are prioritized, increasing the number of annotated tomato flower instances to 929 and raising its mAP to 59.3, which confirms the tangible benefit of active learning for minority-class recognition.

When the dataset is further expanded to the full set (2,186 images), the overall mAP for tomatoes and green tomatoes decreases slightly (from 78.9 to 77.7 and from 65.1 to 64.3, respectively). A likely reason is that many of the additional images not selected in earlier AL rounds are more challenging—e.g., heavily occluded or blurred—thereby lowering the average performance. This illustrates that increasing data quantity alone does not guarantee accuracy improvements.

In contrast, the tomato flower class maintains nearly the same performance on the full dataset (59.3 → 59.4), suggesting that active learning has already driven the model to focus on high-information, representative minority-class samples in earlier stages, leading to more stable feature representations and reducing the impact of subsequent imbalance.

In summary, active learning not only improves overall data utilization but also substantially enhances minority-class performance under class-imbalanced conditions. For majority classes, however, indiscriminate late-stage data expansion may introduce a high proportion of difficult samples and slightly degrade overall performance. These findings indicate that, in practical agricultural scenarios, simply annotating large volumes of data is suboptimal; combining active learning with selective data expansion is more effective for maintaining stable and reliable recognition under class imbalance.

**4.8. Comparison with Lightweight YOLO Baselines**

This section presents a comparative analysis of the object detection models listed in Table 4, including YOLOv5n, YOLOv7-tiny, YOLOv10n, YOLOv11n, YOLOv9t, and the proposed LWA-YOLO. The comparison considers detection performance, model parameters, and computational complexity to evaluate the trade-off between accuracy and efficiency, and to assess their suitability for deployment on agricultural edge devices.

YOLOv7-tiny targets real-time inference and high throughput. However, its backbone is only moderately simplified and retains a relatively large number of feature channels to preserve multi-scale information, resulting in 6.02M parameters and 13.2 GFLOPs. Although its $mAP_{50}$ reaches 68.2%, the overall computational cost is relatively high, posing potential risks for stable deployment on resource-constrained edge platforms.

YOLOv10n substantially compresses both parameters (2.7M) and FLOPs (8.2G), achieving good computational efficiency. Yet the aggressive modular and block-level compression limits feature representation capacity: on the multi-scale, small-object–dominated agricultural dataset used in this study, its $mAP_{50}$ is only 43.2%, indicating that excessive compression is detrimental in complex scenes. YOLOv11n alleviates this to some extent by introducing deeper modules and lightweight optimizations, improving mAP$_{50}$ to 64.3%. Nonetheless, its ability to model fine-grained features remains insufficient, and it still struggles with blurred boundaries and highly similar crop categories (e.g., tomatoes vs. tomato flowers), suggesting room for improvement in multi-scale and fine-detail modeling.

YOLOv9t, the lightweight variant of the YOLOv9 family, employs the RepNCSPELAN4 architecture and achieves a relatively good balance between accuracy and efficiency, attaining 68.8% mAP$_{50}$, the highest among the baseline models. However, it does not incorporate specialized attention mechanisms or designs explicitly targeting small-object enhancement. As a result, its multi-scale feature integration is still limited in scenarios with highly confusing samples and large scale variation.

In contrast, the proposed LWA-YOLO maintains an extremely compact model size (1.77M parameters, 7.3 GFLOPs) while using the C3Ghost module to reduce redundant computation and the C2PSA attention mechanism to strengthen joint spatial–channel feature integration. This combination improves recognition of small objects and samples with ambiguous boundaries. In addition, the Mosaic9-based data augmentation strategy further enhances multi-scale learning and generalization. On the experimental dataset, LWA-YOLO achieves 67.8% mAP$_{50}$, offering the best balance between detection performance and computational cost among the compared models.

Overall, YOLOv7-tiny provides reasonable accuracy but at a relatively high computational cost; YOLOv10n and YOLOv11n are limited by over-compression or inadequate feature modeling capacity; and YOLOv9t, while strong, lacks adaptations tailored to agricultural imagery. LWA-YOLO, by contrast, achieves substantially lower parameter count and FLOPs than the other models while maintaining performance comparable to, or better than, most lightweight baselines. This demonstrates that LWA-YOLO is particularly well suited for real-time deployment on resource-limited smart agricultural devices.

| *Model* | *P* | *R* | *mAP$_{50}$* | *Param.* | *FLOPs* |
|---|---|---|---|---|---|
| YOLOv7-tiny[26] | 71.3 | 64.2 | 68.2 | 6.02M | 13.2G |
| YOLOv10n[27] | 70.7 | 63.3 | 67.3 | 2.7M | 8.2G |
| YOLOv11n[22] | 67 | 60.6 | 64.3 | 2.58M | 6.3G |
| YOLOv9t[20] | 70.1 | 63.7 | 68.8 | 2.66M | 10.7G |
| LWA-YOLOv9 (OURS) | 63.3 | 66 | **67.8** | **1.77M** | 7.3G |

Table 4. Model Performance Comparison

### 4.9. Experimental Summary

Section 4.3 conducted ablation studies on the proposed lightweight modules, showing that replacing RepNCSPELAN4 with C3Ghost significantly reduces parameters and FLOPs, and that adding C2PSA restores deep feature representation and improves small-object recognition. Section 4.4 evaluated the effect of the dynamic Mosaic9 strategy, demonstrating that enabling Mosaic9 in early epochs and disabling it later enhances multi-scale feature learning and generalization. Section 4.5 analyzed active learning performance under different uncertainty aggregation schemes, confirming that uncertainty-driven sampling improves data efficiency under limited annotation budgets. Section 4.6 further compared Max, Sum, and Average scoring with Move and Copy update strategies, revealing that a Max-based Move policy yields the most stable and rapid performance gains. Section 4.7 validated the impact of class imbalance, showing that active learning markedly boosts tomato-flower detection despite its minority status. Finally, Section 4.8 compared the proposed LWA-YOLO with other lightweight YOLO baselines, where LWA-YOLO achieved the best trade-off with 67.8% mAP$_{50}$ using only 1.77M parameters and 7.3 GFLOPs, confirming its suitability for edge deployment in agricultural scenarios.

### 5. Conclusion

We proposed an active learning–driven lightweight YOLOv9 framework for real-time detection of tomatoes and tomato flowers in raw greenhouse imagery, combining C3Ghost module, C2PSA attention, Dynamic Mosaic9 augmentation, and confidence-based active learning selection.

On the collected greenhouse dataset, the method achieves 67.8% mAP50 with 1.77M parameters and 7.3 GFLOPs, and reaches near-full-data performance using only 1,730 labeled images via AL, demonstrating practicality for edge-device deployment under limited labeling budgets.

Future work includes improving minority-class acquisition, integrating image-quality filtering to avoid low-quality samples during AL, and validating end-to-end latency/energy on embedded hardware platforms used by agricultural robots.

# Appendix A
**Comparison of Results of Dataset Increase at Each Stage of Active Learning**

| Method | Strategy | Round1 | 2 | 3 | 4 | 5 | 6 | 7 | 8 |
|---|---|---|---|---|---|---|---|---|---|
| Average | Move | 230 | 730 | 1230 | 1730 | 2186 | | | |
| | Copy | 230 | 730 | 980 | 1114 | 1240 | 1315 | 1352 | 1391 |
| Max | Move | 230 | 730 | 1230 | 1730 | 2186 | | | |
| | Copy | 230 | 730 | 1074 | 1318 | 1512 | 1636 | 1730 | 1794 |
| Sum | Move | 230 | 730 | 1230 | 1730 | 2186 | | | |
| | Copy | 230 | 730 | 884 | 1006 | 1056 | 1095 | 1128 | 1154 |

# Appendix B
**Comparison of Data Annotation Quantity under Move Strategy**

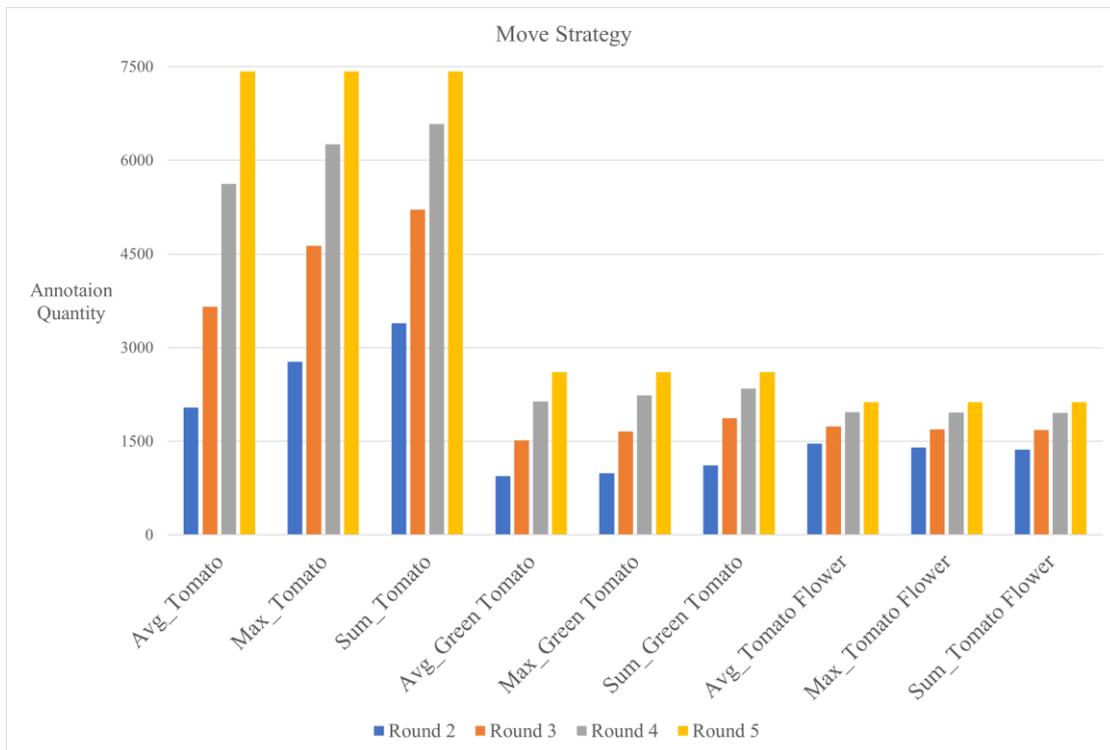

# Appendix C
**Comparison of Data Annotation Quantity under Copy Strategy**

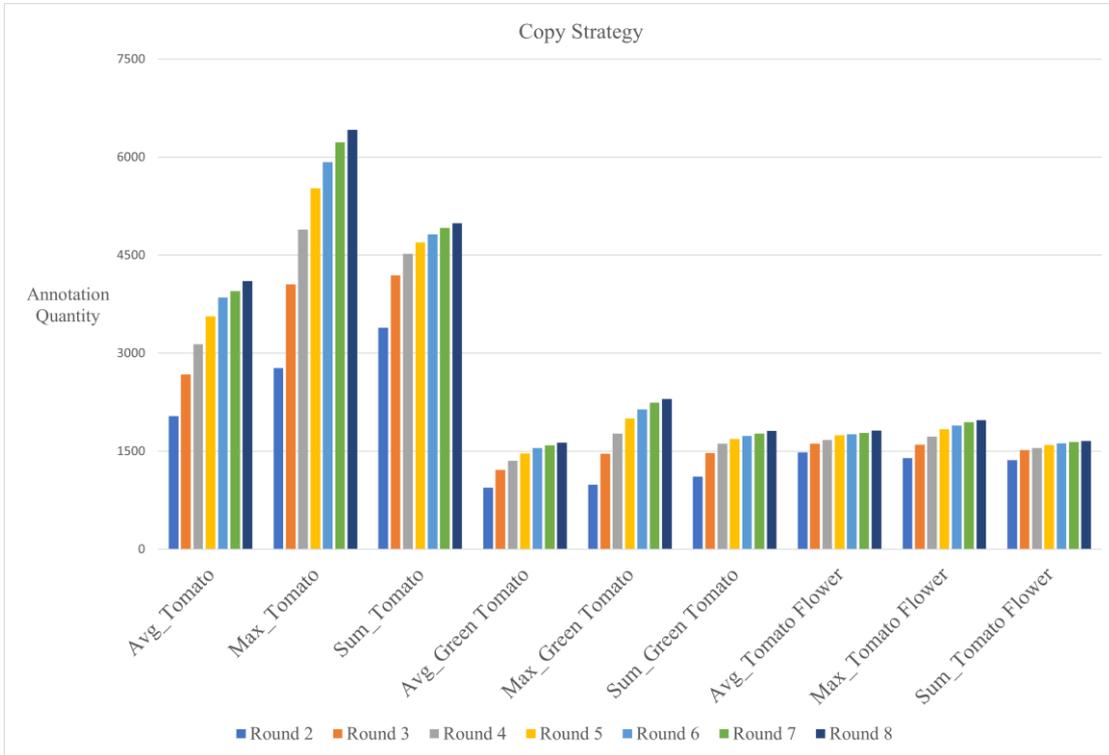

# Appendix D
**Comparison of Training Score Results at Each Stage of Active Learning**

| Method | Strategy | Round1 | 2 | 3 | 4 | 5 | 6 | 7 | 8 |
|---|---|---|---|---|---|---|---|---|---|
| **Average** | Move | 41.2 | 54.9 | 62.7 | 64.9 | 67.1 | | | |
| | Copy | 41.2 | 54.9 | 51.4 | 65.1 | 61.1 | 57.5 | 62.9 | 63.2 |
| **Max** | Move | 41.2 | 55.2 | 62.8 | **67.8** | 67.1 | | | |
| | Copy | 41.2 | 55.2 | 62.1 | 63.1 | 64.6 | 66 | 66.8 | 65 |
| **Sum** | Move | 41.2 | 55.5 | 64.7 | 64.6 | 67.1 | | | |
| | Copy | 41.2 | 55.5 | 52.4 | 58.4 | 61.3 | 58.7 | 62.5 | 64.9 |

# Appendix E

**Comparison of the Number of Categories and Scores in the Initial, Full, and Best-performing Datasets.**

|  | mAP | Tomato(mAP/ label numbers) | Green tomato(mAP/ label numbers) | Tomato flower(mAP/ label numbers) |
|---|---|---|---|---|
| Initial (230) | 41.2 | 65.4 / 742 | 26.8 / 291 | 21.1 / 110 |
| Active Learning (1730) | 67.8 | 78.9 / 6256 | 65.1 / 2232 | 59.3 / 929 |
| Full (2186) | 67.1 | 77.7 / 7424 | 64.3 / 2607 | 59.4 / 1091 |